\documentclass{article}
\usepackage{spconf,amsmath,graphicx}

\usepackage{spconf,amsmath,booktabs,amssymb,graphicx,subfigure,threeparttable,multirow,float}
\usepackage{hyperref}
\hypersetup{
    colorlinks=true,
    linkcolor=red,
    filecolor=red,      
    urlcolor=red,
    citecolor=green,
}

\title{WASE: LEARNING WHEN TO ATTEND FOR SPEAKER EXTRACTION\\ IN COCKTAIL PARTY ENVIRONMENTS}
\name{Yunzhe Hao$^1$$^,$$^2$, Jiaming Xu$^1$$^,$$^3$$^\dagger$\thanks{$^\dagger$Corresponding author}, Peng Zhang$^1$$^,$$^3$, Bo Xu$^1$$^,$$^2$$^,$$^3$$^,$$^4$$^\dagger$}
\address{
$^1$Institute of Automation, Chinese Academy of Sciences (CASIA), Beijing, China\\
$^2$School of Future Technology, University of Chinese Academy of Sciences, Beijing, China\\
$^3$School of Artificial Intelligence, University of Chinese Academy of Sciences, Beijing, China\\
$^4$Center for Excellence in Brain Science and Intelligence Technology, CAS, Beijing, China}

\begin{document}
\ninept
\maketitle
\begin{abstract}
In the speaker extraction problem, it is found that additional information from the target speaker contributes to the tracking and extraction of the target speaker, which includes voiceprint, lip movement, facial expression, and spatial information. However, no one cares for the cue of sound onset, which has been emphasized in the auditory scene analysis and psychology. Inspired by it, we explicitly modeled the onset cue and verified the effectiveness in the speaker extraction task. We further extended to the onset/offset cues and got performance improvement. From the perspective of tasks, our onset/offset-based model completes the composite task, a complementary combination of speaker extraction and speaker-dependent voice activity detection. We also combined voiceprint with onset/offset cues. Voiceprint models voice characteristics of the target while onset/offset models the start/end information of the speech. From the perspective of auditory scene analysis, the combination of two perception cues can promote the integrity of the auditory object. The experiment results are also close to state-of-the-art performance, using nearly half of the parameters. We hope that this work will inspire communities of speech processing and psychology, and contribute to communication between them. Our code will be available in \url{https://github.com/aispeech-lab/wase/}.

\end{abstract}
\begin{keywords}
onset cue, onset/offset cues, voiceprint, speaker extraction, cocktail party problem
\end{keywords}
\section{Introduction}
\label{sec:intro}

The cocktail party effect is the phenomenon that the brain focuses one’s auditory attention on specific stimuli while filtering out other stimuli \cite{getzmann2017switching, bronkhorst2000cocktail}. For example, in complex auditory scenes, people can pay attention to the chat partner while ignoring other people or noise in the background \cite{bronkhorst2000cocktail,cherry1953some}. Researchers in different fields, which include psychology, neuroscience, and information science, have made many efforts to analyze and model the brain’s ability of auditory attention \cite{bregman1994auditory,winkler2009modeling,nelken2008processing}. In the artificial intelligence community, with the rapid development of deep learning, more and more methods have been proposed to model the cocktail party effect. According to the formal definition of the problem, these models can be mainly divided into two types, speech separation and speaker extraction \cite{Wang2018c,michelsanti2020overview}. Suppose that there are $N$ sources, $s_1(t)$, $s_2(t)$, …, $s_N(t)$, then the mixture signal $x(t) \in \mathbb{R}^{1 \times T}$ of $N$ sources can be defined as:

\begin{equation}
  \label{eq:mixture}
  x(t)=\sum_{i=1}^{N}{s_i(t)}.
\end{equation}

\begin{figure}[tbp]
  \centering
  \includegraphics[width=\linewidth]{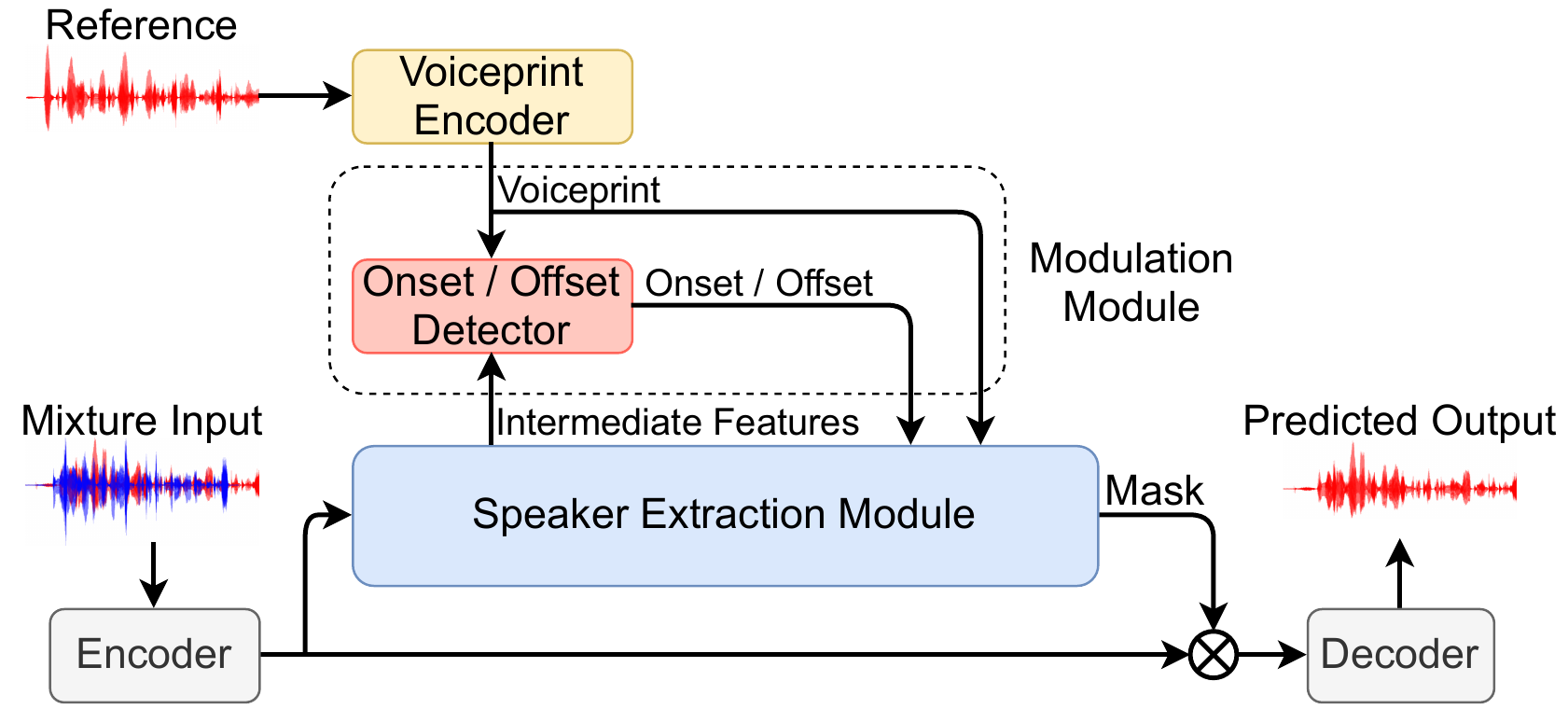}
  \caption{The framework of WASE. Note that offset cue and voiceprint cue are optional to injected into speaker extraction module.}
  \label{fig:framework}
\end{figure}

The problem of speech separation can be formulated in terms of estimating $N$ sources, while speaker extraction estimating only one source. Since the speech separation aims to extract each source, this leads to two problems, uncertain number of source problem and label permutation problem \cite{yu2017permutation}. Specifically, $N$ may change dynamically, even unknown sometimes, which result in the uncertain number of source problem; the label permutation problem is that how to add supervision signals to correct channels and avoid training failure caused by gradient conflict. Researchers proposed some solutions represented by deep clustering and permutation invariant training \cite{yu2017permutation,hershey2016deep}. In comparison, speaker extraction takes a different path from the prior. Speaker extraction utilizes additional information that can identify the speaker of interest and regards the information as cues to extract the target speaker. Because speaker extraction is only interested in one speaker, it avoids the two problems above. The additional information includes voiceprint (pitch, identity), visual information (lip movement, facial expression), spatial information (azimuth) \cite{michelsanti2020overview,wang2019voicefilter,afouras2019my,gu2019neural}.

Except the above-mentioned cues, the onset signal is also an important cue in the auditory scene analysis \cite{shamma2011temporal,szabo2016computational}. Previous psychological studies have found that infants are more sensitive to the synchronization between the onset times of auditory and visual stimuli, instead of continuous temporal coherence between auditory and visual events \cite{lewkowicz2010infant,desjardins2004integration}. Infants, who have little experience in language, cannot utilize additional temporal and phonetic information provided by visual signal. However, they can take advantage of the onset cue and benefit from when to hear \cite{werner2009effects}.

In this paper, we propose a novel model that can learn When to Attend for Speaker Extraction (dubbed WASE). WASE can detect the onset signal and regard it as important guidance, as shown in Fig.~\ref{fig:framework}. Specifically, we first obtain the speaker voiceprint from the reference voice. And then we use the onset detector to predict the onset of noisy voice intermediate features with the help of speaker voiceprint. Finally onset signal directly filters intermediate features, which inhibits the signal before onset and guides the extraction module to extract the target speaker from a specific onset time.
We propose a multi-task learning strategy to support our model and use a unified training objective that considers both speaker extraction and onset detection tasks. The oracle onset is generated from the voice activity detection (VAD) results of the clean voice. The training objective of onset detection is to minimize the cross-entropy (CE) loss between oracle onset and the predicted onset in intermediate feature space.
We extend onset cue to onset/offset cues, which means not only the start of voice but also the end of voice is predicted. Interestingly, our onset/offset prediction task is actually speaker-dependent voice activity detection (SDVAD) \cite{chen2020end}. Therefore, the onset/offset-based model completes a combination task of speaker extraction and SDVAD.
Cross fusion of multiple cues is an important mechanism of the brain to process related information from complex scenes and form auditory objects. Therefore, we also extend a single cue to multiple cues, i.e., combine onset/offset cues with voiceprint cue.

Our contributions can be concluded as follow:

\begin{enumerate}
\item As far as we know, it is the first time to explicitly model onset cue in speaker extraction problem, which is considered an important cue for the cocktail party effect in both fields of auditory scene analysis and psychology;

\item We extend the onset cue to onset/offset cues. Since our model reconstructs clean voice and predicts the start/end of voice simultaneously, it completes the speaker extraction task and SDVAD task, which have complementary advantages;

\item We further combine the onset/offset cues with the voiceprint cue. The onset/offset cues play a role in the time dimension while the voiceprint plays a role in the feature dimension, respectively, which is beneficial for the integrity of the auditory object.
\end{enumerate}

\section{RELATION TO PRIOR WORK}
\label{sec:related}
Several prior works for speaker extraction have studied various cues about the target speaker, such as voiceprint \cite{wang2019voicefilter,Xu2020SpEx,koizumi2020speech}, lip movement \cite{afouras2019my,Ephrat2018Looking}, facial appearance \cite{chung2020facefilter}, and spatial information \cite{gu2019neural}. Voiceprint models voice characteristics. Facial appearance models the cross-modal relationship between facial appearance and the voice characteristics of the target speaker. Lip movement information has strong coherence with the dynamic changes of voice, which can enhance the voice signal. The spatial information contained in multi-channel acoustic recordings can filter the speaker in a specific location. Among the cues mentioned above, although lip movement contains onset/offset cues, no study has explicitly modeled it. Therefore, our model is the first time to directly focus on the role of onset/offset cues in speaker extraction.

\section{METHOD}
\label{sec:method}
\subsection{Onset Cue}
\label{ssec:onset}
Onset cue is one of the most important cues in the auditory scene analysis. We take advantage of the reference voice to obtain the onset cue from the mixture input. The reference voice can be represented by $r(t) \in \mathbb{R}^{1 \times {T^r}}$, where $T^r$ denotes the length of the reference voice. $r(t)$ is transformed into a $C$-dimensional representation, ${v \in \mathbb{R}^{C \times 1}}$, by voiceprint encoder. The intermediate features of the mixture can be represented by $U \in \mathbb{R}^{C \times L}$, where $L$ denotes the length of intermediate features. The onset detector searches for the onset signal in $U$ with the guidance of $v$. The onset cue can be defined as:
\begin{equation}
  \label{eq:detector}
  o={\mathcal H}(U, v),
\end{equation}
where $\mathcal H(\cdot, \cdot)$ is the onset cue detector, as shown in Fig.~\ref{fig:model}(C-b). $o \in \mathbb{R}^{1 \times L}$ is a 1-dimensional (1-D) vector of the same length with $U$, and the value of each element in vector represents the current position relationship with onset point, i.e., the value larger than threshold denotes after onset while smaller than threshold denotes before onset. The range of value is $[0, 1]$ and the threshold is $0.5$. That is, the single change point across the threshold is the most important information contained in $o$. To offer guidance in extracting the target speaker voice, $o$ is injected to the intermediate features as follow:
\begin{equation}
  \label{eq:onset_inject}
  U'=U \odot o,
\end{equation}
where $\odot$ denotes element-wise multiplication. $o$ is supervised by oracle onset, which is generated from the clean voice. Specifically, the start time of clean voice is detected by VAD, and we use it to generate a binary vector, ${b \in \mathbb{R}^{1 \times T}}$, with a single change point, i.e., $0$ is before the start time and $1$ after it. But there is a problem with the length of the generated binary vector and $o$ is $T$ and $L$, respectively. To make length equal, we downsample the binary vector with the fixed stride, which is equal to the length of the encoder stride. The downsampled results are named oracle onset in this paper. Note that the stride size of the encoder is $1$ ms in our model, which seems short enough and can hardly influence the precision of oracle onset.

\subsection{Extensions of Onset Cue}
\label{ssec:extensions}
\begin{figure*}[t]
  \centering
  \includegraphics[scale=0.37]{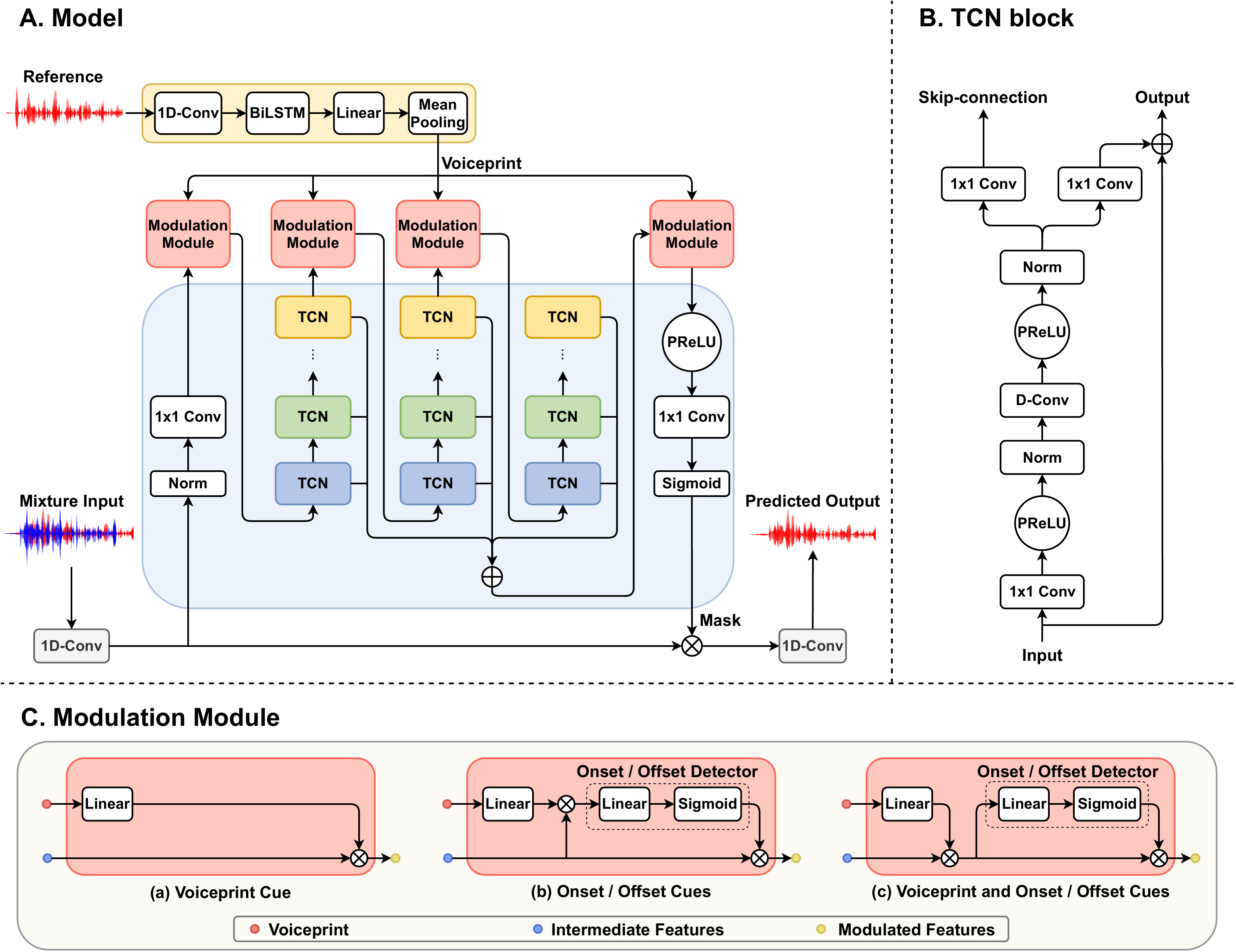}
  \caption{An illustration of our WASE model. (A): The overall architecture of the proposed WASE. (B): Details of TCN block in speaker extraction module. `D-Conv' indicates a depthwise convolution. (C): Three modulated methods based on different cues.}
  \label{fig:model}
\end{figure*}

\noindent{\bf{Onset/Offset Cues:}}
Intuitively, offset cue can bring additional benefits. Therefore, we extend the onset cue to onset/offset cues. The structure of the onset/offset detector is the same as the onset detector, as shown in Fig.~\ref{fig:model}(C-b). The only thing that changed is the supervision signal. Specifically, we use both the start time and the end time of clean voice to generate $b$, which has two change points. Then we downsample $b$ to obtain oracle onset/offset cues.

\noindent{\bf{Voiceprint Cues:}}
Onset/offset cues filter mixture intermediate features in the time dimension, while voiceprint filters in the feature dimension. Therefore, we manage to take advantage of both cues in one model. Specifically, the voiceprint cue first filter the noisy voice, and then onset/offset cues filter the noisy voice, as shown in Fig.~\ref{fig:model}(C-c). In contrast, we also show the only voiceprint-based modulation method in Fig.~\ref{fig:model}(C-a).

\subsection{Model Structure}
\label{ssec:model}
The onset cue and its extensions have been described above. Here we introduce the base model in detail. The structure of the base model is adapted based on our previous proposed framework \cite{hao2020unified}, which includes five modules: voiceprint encoder, onset/offset detector, speech encoder, speech decoder, and speaker extraction module, as shown in Fig.~\ref{fig:model}(A). The model has both the reference voice of the target speaker and the noisy voice as inputs. It is optimized by a multi-task learning strategy, which maximizes the training objective of speaker extraction, and minimizes the loss between the predicted onset cue and the oracle onset cue, simultaneously. The main difference between the current model and the previous model lies in two points: 1) we introduce a more complex module to model more cues, instead of a simple voiceprint cue; 2) we replace the previous feature extraction method of reference voice, short-time Fourier transform (STFT), with the time-coding method, i.e., convolution layer.

\noindent{\bf{Voiceprint Encoder:}}
We obtain the reference voice feature by the time-coding method with a 1-D convolutional layer, whose kernel size is $256$ and stride size is $64$, i.e., $32$ ms and $8$ ms in $8$ kHz. The reference voice feature is processed using two-layer bidirectional long short-term memory (LSTM) and one fully-connected layer. Moreover, we use a mean-pooling to squeeze the time dimension of the reference voice feature and finally get the voiceprint vector $v$. Note that the voiceprint encoder weights will be frozen after $150$ epochs to avoid overfitting.

\noindent{\bf{Speech Encoder \& Decoder:}}
The encoder and decoder follow the same structure in \cite{luo2019conv}, which use a 1-D convolutional layer and a 1-D transpose convolutional layer without bias, respectively. The time-coding method has some advantages over the traditional STFT method, including no phase reconstruction problem, trainable weights, and finer coding granularity.

\noindent{\bf{Extraction Module:}}
We use the temporal convolutional network (TCN) block as the main structure in the speaker extraction module \cite{luo2019conv,Lea2016a}. As shown in Fig.~\ref{fig:model}(B), the TCN block contains pointwise convolution ($1\times1$ Conv), parametric rectified linear unit (PReLU), normalization (Norm), and depthwise convolution (D-Conv). The TCN block has two output paths, one of which is fed to the next TCN block and the other is summarized to obtain the final mask. TCN network has three groups, and each group has eight TCN blocks. The dilation factor of D-Conv increases exponentially by $2$ in each group, which make convolutional receptive fields enlarge layer-by-layer and allow the extraction module to capture correlation in a longer time scale.

We multiply specific cues (onset/offset, voiceprint) with the intermediate features of the speaker extraction module, and inject the results back to guide speaker extraction, as shown in Fig.~\ref{fig:model}(C). 

\noindent{\bf{Loss Function:}}
We use the multi-task learning strategy to train our model. One task is reconstructing the target speaker voice, which is supervised by the scale-invariant source-to-noise ratio (SI-SNR) between predicted wave $\hat{s}$ and clean wave $s$. The SI-SNR can be defined as:
\begin{align}
  \label{eq:SI-SNR}
  \begin{cases}
  s_{target}=\frac{\left \langle {\hat{s},s} \right \rangle s}{{\left \| s \right \|}^2};\\
  e_{noise}=\hat{s}-s_{target};\\
  SI-SNR=10 \log_{10} {\frac{{\left \| s_{target} \right \|}^2}{{\left \| e_{noise} \right \|}^2}}, \\  
  \end{cases}
\end{align}%
where ${\left \langle {s,s} \right \rangle}$ and ${{\left \| s \right \|}^2}$ are the signal power. The other task is the onset/offset detection, which is supervised by CE loss between the predicted onset/offset vector and the oracle onset/offset. The ratio between two losses is 1.

\section{Experiment Settings}
\label{sec:setting}
We performed experiments on speech separation and speaker extraction benchmark dataset WSJ0-2mix \cite{hershey2016deep}. The training set contains $101$ speakers, and the test set includes $18$ unseen speakers. We downsampled all samples to $8$ kHz. For training samples, we selected randomly two speakers and regarded one as the target speaker and the other as the interfering speaker. Then we sampled two voices from the target and regarded them as clean voice and reference voice, respectively. We sampled one voice from the interfering speaker and set it as the interfering voice. The noisy voice is generated by mixing clean voice and interfering voice at random signal-to-noise ratios (SNR) between $-2.5$ dB and $2.5$ dB, and the length is limited to $4$ seconds. Note that we also used interfering voice as a supervision signal to guide interfering voice prediction in the training phase, which is generated by decoding noisy intermediate features masked by residual of the target mask. In the experiments of onset/offset, we pad silence with a random length of $[200, 800]$ ms at the end of the target/interfering voice, which ensure the existence of offset. In the test phase, we modified the test dataset to match the speaker extraction task. Specifically, we set each one of the two speakers in the original test dataset as the target in turn. We obtained $6000$ samples finally, which is twice the number of original test samples.

We use Adam \cite{kingma2014adam} optimizer with an initial learning rate of $1e^{-3}$, decreasing by half if no improvement in recent $10$ epochs. Our model is trained until the performance of the evaluation set is not improved in $10$ consecutive epochs after the learning rate equals $2.5 \cdot e^{-4}$. We built our model in Pytorch.

\section{Results}
\label{sec:results}
To illustrate the effectiveness of the onset/offset cues, we first trained the base model with given oracle onset/offset cues. As shown in Tab.~\ref{tab:oracle cue}, both models achieved good performance, which showed that the onset/offset guidance information is enough to extract the target speaker. It is reasonable that the onset/offset-based model is better than the onset-based model because of more guidance information in onset-offset cues. And then we used the modulated module to predict onset/offset cues and reported metrics of speaker extraction and SDVAD. The signal-to-distortion ratio (SDR) improvements of models with predicted onset/offset cues is not as good as responding oracle onset/offset cues. However, the model based on the predicted onset cue is better than the oracle onset. We analyze this phenomenon and find that the learned onset cue contains some behavior similar to offset cue at the end of target speech, especially when the target speech is much shorter than the interfering speech. In other words, the network spontaneously learns behaviors similar to onset/offset cues. Therefore, onset-only cue leads to better results than oracle onset. We reported the SDVAD results for the frame-level evaluation in terms of accuracy (ACC) and F-score (F1, the harmonic mean of precision and recall). The ACC and F1 results show that our model could capture the target speaker onset/offset signal. The higher layer, the higher ACC and F1 performance, which is more obvious in onset/offset. It indicates that the onset/offset cues of the target speaker are abstract representations, and it is difficult to effective detection only relying on the voiceprint and the representation of low-layer in our model. The visual information may be a better choice for onset/offset detection solely relying on low-layer representation, e.g., lip movement.

\begin{table}[htbp]
  \caption{SDR improvements (dB) with onset/offset cues on WSJ0-2mix dataset.}
  \label{tab:oracle cue}
  \centering
  \begin{tabular}{llll}
    \toprule
    \textbf{Cues used} & \textbf{ACC(\%)} & \textbf{F1(\%)} & \textbf{SDRi(dB)} \\
    \midrule
    oracle onset & - & - & 14.78 \\
    oracle onset / offset & - & - & 17.02 \\
    \midrule
    onset & 91/90/92/87 & 95/94/95/92 & 16.31 \\
    onset / offset & 66/90/93/96 & 78/92/94/97 & 16.75 \\
    \bottomrule
  \end{tabular}
\end{table}

The voiceprint representation is difficult to distinguish people with similar voices, while onset/offset cues are challenging to handle simultaneous voices. The model with all mentioned cues above could combine the advantages of all cues and form a more robust representation of the auditory object. In Tab.~\ref{tab:voiceprint}, compared with the model based on single voiceprint cue, the performance of the model with two cues is improved while the additional parameters are negligible.

\begin{table}[htbp]
  \caption{SDR improvements (dB) with two cues on WSJ0-2mix dataset.}
  \label{tab:voiceprint}
  \centering
  \begin{tabular}{llll}
    \toprule
    \textbf{Cues used} & \textbf{\#Params} & \textbf{SDR(dB)} & \textbf{SDRi(dB)} \\
    \midrule
    voiceprint & 7.5M & 16.27 & 16.14 \\
    onset + voiceprint & 7.5M & 17.12 & 16.99 \\
    onset / offset + voiceprint & 7.5M & 17.18 & 17.05 \\
    \bottomrule
  \end{tabular}
\end{table}

A comparison with the previously published results on the WSJ0-2mix dataset is shown in Tab.~\ref{tab:others}. Although our best performance is slightly worse than SpEx+, our parameter quantity is nearly half of it, which shows the potential of onset/offset cues and multi-cue integration.
\begin{table}[htbp]
  \caption{SDR improvements (dB) with different speaker extraction methods based on WSJ0-2mix dataset.}
  \label{tab:others}
  \centering
  \begin{tabular}{lllll}
    \toprule
    \textbf{Methods} & \textbf{\#Params} & \textbf{SDRi(dB)} \\
    \midrule
    SBF-MTSAL~\cite{Chenglin2019Optimization} & 19.3M & 7.30 \\
    SBF-MTSAL-Concat~\cite{Chenglin2019Optimization} & 8.9M & 8.39 \\
    SpEx~\cite{Xu2020SpEx} & 10.8M & 14.6 \\
    SpEx+~\cite{xu2020spex_plus} & 13.3M & {\bf 17.2} \\
    WASE (onset / offset + voiceprint) & 7.5M & 17.05 \\
    \bottomrule
  \end{tabular}
\end{table}

\section{CONCLUSION}
\label{sec:conclusion}
Inspired by auditory scene analysis and psychology, we focus on the onset cue and verify the effectiveness in speaker extraction. Then we explicitly model onset cue using reference voice and obtain comparable performance on the benchmark dataset. We extend onset cue to onset/offset cues, and the performance further improved. Since the onset/offset-based model completes the combination tasks of speaker extraction and speaker-aware voice activity detection, it verifies the mutual benefits between the two related tasks. We also combine the onset/offset cues and voiceprint cue. Onset/offset cues model start/end time of voice and voiceprint cue models the voice characteristics. Therefore, the combination of these perceptual cues can promote the integrity of auditory objects. The experimental performance is also further improved.

\section{Acknowledgments}
This work was supported by the Major Project for New Generation of AI (Grant No. 2018AAA0100400), and the Strategic Priority Research Program of the Chinese Academy of Sciences (Grant No. XDB32070000).

\vfill\pagebreak

\bibliographystyle{IEEEbib}
\bibliography{strings,refs}

\end{document}